\documentclass[useAMS]{mn2e}

\usepackage{times}
\usepackage{graphicx}
\usepackage{amsmath}
\usepackage{amssymb}
\usepackage{color}
\newcommand{\beq}{\begin{equation}}
\newcommand{\bea}{\begin{eqnarray}}
\newcommand{\eeq}{\end{equation}}
\newcommand{\eea}{\end{eqnarray}}

\title[The magnetization of gamma-ray bursts afterglows]{On the
  magnetization of gamma-ray burst blast waves}

\author[M. Lemoine, Z. Li \& X.-Y. Wang] {Martin
  Lemoine$^1$\thanks{e-mail:{\tt lemoine@iap.fr}}, Zhuo
  Li$^{2,3}$\thanks{e-mail:{\tt zhuo.li@pku.edu.cn}},
  Xiang-Yu Wang$^{4,5}$\thanks{e-mail:{\tt xywang@nju.edu.cn}} \\
  $^1$Institut d'Astrophysique de Paris, CNRS, UPMC,
  98 bis boulevard Arago, F-75014 Paris, France\\
  $^2$ Department of Astronomy / Kavli Institute for Astronomy and
  Astrophysics, Peking
  University,  Beijing 100871, China\\
  $^3$ Key Laboratory for the Structure and Evolution of Celestial
  Objects, Chinese Academy of Sciences, Kunming 650011, China\\
  $^4$School of Astronomy and Space Science, Nanjing University,
  Nanjing, 210093, China\\
  $^5$ Key laboratory of Modern Astronomy and Astrophysics (Nanjing
  University), Ministry of Education, Nanjing 210093, China }

\begin{document}

\date{}

%\pagerange{\pageref{firstpage}--\pageref{lastpage}} 
\pubyear{2008}

\maketitle

\label{firstpage}

\begin{abstract}
  The origin of magnetic fields that permeate the blast waves of
  gamma-ray bursts (GRBs) is a long-standing problem.  The present
  paper argues that in four GRBs revealing extended emission at
  $>100\,{\rm MeV}$, with follow-up in the radio, optical and X-ray
  domains at later times, this magnetization can be described as the
  partial decay of the micro-turbulence that is generated in the shock
  precursor. Assuming that the bulk of the extended emission
  $>100\,$MeV can be interpreted as synchrotron emission of
  shock-accelerated electrons, we model the multi-wavelength light
  curves of GRB~090902B, GRB~090323, GRB~090328 and GRB~110731A, using
  a simplified then a full synchrotron calculation with
  power-law-decaying micro-turbulence $\epsilon_B
  \,\propto\,t^{\alpha_t}$ ($t$ denotes the time since injection
  through the shock, in the comoving blast frame). We find that these
  models point to a consistent value of the decay exponent
  $-0.5\,\lesssim\,\alpha_t\,\lesssim\, -0.4$.
\end{abstract}

\begin{keywords} 
acceleration of particles -- shock waves -- gamma-ray bursts: general
\end{keywords}

\section{Introduction}\label{sec:introd}
In principle, the multi-wavelength light curves of gamma-ray bursts
(GRBs) in the afterglow phase open a remarkable window on the physics
of relativistic, weakly magnetized collisionless shock waves: these
light curves are indeed thought to result from the synchrotron process
of electrons accelerated at the external shock wave, so that their
modelling leads to two microphysical parameters of importance: the
fraction of shock dissipated energy stored in the suprathermal
electron population, $\epsilon_e$, and in the self-generated
electromagnetic turbulence, $\epsilon_B$.

From a theoretical point of view, one expects $\epsilon_B\,\sim\, 0.1$
at the shock front (and $\epsilon_e\,\sim\,0.1$): the shock wave forms
when a magnetic barrier on the ion skin depth scale
$\,\sim\,c/\omega_{\rm pi}$ builds up through small-scale
electromagnetic instabilities, up to the level at which it can deflect
by an angle of the order of unity the incoming particles, which carry
Lorentz factor $\gamma_{\rm sh}$ in the shock front frame; this
demands $\epsilon_B\,\sim\,1/4$. This picture has been validated by
high performance particle-in-cell (PIC) simulations (e.g. Spitkovsky
2008, Martins et al. 2009, Haugb\o lle 2011, Sironi \& Spitkovsky
2011, 2013), and supported by theoretical arguments (e.g. Medvedev \&
Loeb 1999).  However, on such small plasma scales, the
micro-turbulence should decay rapidly behind the shock (e.g. Gruzinov
\& Waxman 1999), whereas early afterglow models of GRBs have pointed
to finite, substantial values of $\epsilon_B$ on the (comoving) scale
of the blast $\sim c t_{\rm dyn}$ [with $t_{\rm
  dyn}\,\sim\,r/(\gamma_{\rm b}c)$ the dynamical time-scale,
$\gamma_{\rm b}$ the blast Lorentz factor], many orders of magnitude
larger than the skin depth scale (e.g. Piran 2004, and references
therein)\footnote{We use the standard notation $Q_x\,\equiv\, Q/10^x$
  in CGS units, unless otherwise noted.}: $t_{\rm dyn}\omega_{\rm
  pi}\,\sim\,2\times10^7\,E_{54}^{1/8}n_0^{3/8}
t_{2}^{5/8}$. Nevertheless, the decay of Weibel turbulence has been
observed in dedicated numerical experiments (Chang et al. 2008, Keshet
et al. 2009, Medvedev et al. 2011), although admittedly, such
simulations can probe only a small fraction of a GRB dynamical
time-scale.

The detection of extended high energy emission $>100\,$MeV by the {\it
  Fermi}-LAT instrument in several GRBs has brought in new constraints
in this picture. Most notably, the synchrotron model of this emission
has pointed to values of $\epsilon_B$ much smaller than unity in an
adiabatic scenario (Kumar \& Barniol-Duran 2009, 2010, Barniol-Duran
\& Kumar 2011, He et al. 2011, Liu \& Wang 2011). Kumar \&
Barniol-Duran (2009) have noted that the magnetic field in which the
electrons radiate corresponds to a strength $\sim 10\,\mu$G in the
upstream frame, before shock compression; they therefore interpret
this magnetic field as the simple shock compression of the
interstellar field.  However, the fact that the inferred $\epsilon_B$
lies a few orders of magnitude above the interstellar magnetization
level $\sim10^{-9}$ rather suggests that the electrons radiate in a
partially decayed micro-turbulence (Lemoine 2013); theoretically, such
a picture could reconcile the results of PIC simulations with the
observational determinations of $\epsilon_B$.

In the present work, we push forward this idea and put it to the test
by considering the multi-wavelength light curves of four GRBs observed
in radio, optical, X-ray and GeV in the framework of a decaying
micro-turbulence afterglow scenario. We show that these four bursts
point to a consistent value of the decay index
$-0.5\,\lesssim\,\alpha_t\,\lesssim\,-0.4$, if one assumes that
$\epsilon_B \,\propto\, t^{\alpha_t}$, with $t$ the time since
injection of the plasma through the shock, as measured in the comoving
blast frame, and $\epsilon_B \,\sim\,0.01$ at $t\,=\,100\omega_{\rm
  pi}^{-1}$, as observed in PIC simulations. To do so, we first
present a simplified model of this afterglow (Section~\ref{sec:simp}),
with two radiating zones, in each of which one can use the standard
afterglow model (e.g. Sari et al. 1998); then we provide a detailed
treatment of the power-law decay of the micro-turbulence
(Section~\ref{sec:dec}), improving on Lemoine (2013).  We confront our
findings to previous results in Section~\ref{sec:disc}.

\section{Afterglow model}\label{sec:simp}
\subsection{General considerations}
The calculation of the synchrotron spectrum of a relativistic blast
wave with decaying micro-turbulence can be approximated (and much
simplified) by noting that photons in different frequency bands have
been emitted by electrons of different Lorentz factors, which cool at
different times since their injection, hence in regions of different
magnetic field strengths. In this approximate treatment, one can
therefore use the standard homogeneous afterglow model for each
frequency band, allowing for a possibly different $\epsilon_B$ in each
band. When compared to the detailed calculations with decaying
micro-turbulence, one finds that the above provides a reasonable
approximation, provided the decay index $\alpha_t\,\gtrsim\,-1$. We
make this approximation in the present work and justify it a
posteriori.

According to the above picture, one should take a similar
$\epsilon_{B-}$ for all frequencies $\nu\,<\,\nu_{\rm c}$ that
correspond to Lorentz factors $\gamma\,<\,\gamma_{\rm c}$ such that
the cooling time-scale exceeds the dynamical time-scale, $t_{\rm
  cool}(\gamma)\,\gtrsim\,t_{\rm dyn}$. Such particles indeed radiate
most of their synchrotron energy in the same region, at the back of
the blast. For GRB afterglows with extended $>100\,{\rm MeV}$
emission, in which we are interested here, this concerns the radio and
optical range, and possibly the X-ray range at late times. For those
frequencies, one can therefore use the standard homogeneous
approximation of slowly cooling particles for the calculation of
$F_\nu$.

At the other extreme, GeV photons are likely produced in a region of
strong $\epsilon_B$, due to the short cooling time-scale of the
emitting parent electrons. The large Lorentz factors also generally
imply that inverse Compton losses are negligible in this frequency
range due to Klein-Nishina (KN) suppression, although this should be
verified on a case-to-case basis. Given these assumptions, the
expected flux depends on the ejecta kinetic energy $E$ and
$\epsilon_e$, but very little on the other parameters, $\epsilon_B$ in
particular. Indeed, the energy radiated in the GeV range corresponds
to $\sim (\gamma_8/\gamma_{\rm min})^{2-s}$ times the blast energy
stored in the electron distribution $\propto\,\epsilon_e E$, where
$\gamma_8$ denotes the minimum Lorentz factor of electrons radiating
at $>100\,$MeV. It is easy to see that $\gamma_8/\gamma_{\rm
  min}\,\propto\, \epsilon_B^{-1/4}$, so that the residual dependence
of $F_\nu(>100\,{\rm MeV})$ on $\epsilon_B$ is quite small. As inverse
Compton losses can be neglected at those energies, the flux does not
depend either on the external density $n$. As already noted in Kumar
\& Barniol-Duran (2009), the flux density $F_\nu(>100\,{\rm MeV})$
provides a unique constraint on the model parameters, all the more so
in the present case of decaying micro-turbulence.

The application of the above simple algorithm allows us to evaluate
the parameters of the afterglow in the framework of the standard
model. One outcome of this analysis is the measurement of
$\epsilon_{B-}$, which represents the value of $\epsilon_B$ at the
back of the blast, through the modelling of the radio, optical and
X-ray flux. Since the dynamical time-scale is determined by the
standard parameters of the blast, one can constrain directly the
exponent of power law decay $\alpha_t$:
\begin{equation}
  \alpha_t\,=\, \frac{\log\left[\epsilon_{B-}/\epsilon_{B+}\right]}
  {\log\left[ t_{\rm dyn}/\tau_{\delta
        B}\right]}\ ,
\end{equation}
up to logarithmic corrections dependent on $\tau_{\delta
  B}\,\sim\,100\omega_{\rm pi}^{-1}$, the time scale beyond which
turbulence starts to decay and $\epsilon_{B+}\,\sim\, 0.01$, the value of
the micro-turbulence close to the shock front, both of which are
constrained by PIC simulations.

Care must be taken in the course of this exercise, because for low
$\epsilon_{B-}$, the Compton parameter at the cooling frequency
$Y_{\rm c}\,\gg\,1$, and KN suppression of the inverse Compton process
may be efficient in the X-ray range at late times. The magnitude of KN
suppression at frequency $\nu$ can be quantified through the following
equation:
\begin{eqnarray}
  \Upsilon_{\rm KN}(\nu)&\,\equiv\,&\frac{h\nu_{\rm c}(1+z)}{\gamma_{\rm
      b}}\frac{\gamma(\nu)}{m_e c^2}\nonumber\\
&\,\simeq\,& 50\,
  E_{54}^{1/4}t_{5}^{1.36}A_{35}^{-2.10}\epsilon_{B-,-5}^{-0.80}
\epsilon_{e,-1}^{-1.16}\nu_{17.38}^{1/2}\
  ,
\end{eqnarray}
where $\gamma(\nu)$ denotes the Lorentz factor of electrons whose
(observer frame) synchrotron peak frequency equals $\nu$. For the
numerical values, we have assumed a wind profile of external density
$n=10^{35}A_{35}\,r^{-2}\, $cm$^{-3}$, an electron spectral index
$p=2.2$, $\nu\,>\,\nu_{\rm c}$ with $Y_{\rm c}$ given by Sari \& Esin
(2001) in the slow cooling regime, and $z=1$.  $\Upsilon_{\rm
  KN}\,>\,1$ at X-ray frequencies means that KN suppression of the
inverse Compton cooling is efficient and cannot be ignored.

The optical and radio data of the following light curves always lie
below $\nu_{\rm c}$, in which case the Compton parameter does not
depend on the electron Lorentz factor, $Y(\gamma)\,=\,Y_{\rm c}$, the
Compton parameter at $\gamma_{\rm c}$ (or equivalently, $\nu_{\rm
  c}$). In contrast, at GeV energies KN suppression is so efficient
that the Compton parameter $Y_{>100\,\rm MeV}\,\ll\,1$ (e.g. Wang et
al. 2010, Liu \& Wang 2011). Therefore, inverse Compton losses with
substantial KN suppression, which modify the synchrotron spectrum
(e.g. Nakar et al. 2009, Wang et al. 2010), concern only the X-ray
domain at late times.

We therefore proceed as follows. We first search a solution assuming
$\Upsilon_{\rm KN}\,<\,1$ in the X-ray range, with possibly large
$Y_{\rm c}$. We then compute $\Upsilon_{\rm KN}$, and if
$\Upsilon_{\rm KN}\,>\,1$, we look for another solution in which we
take into account the effect of KN suppression in the X-ray domain,
following Li \& Waxman (2006), Nakar et al. (2009) and Wang et
al. (2010). In particular, we solve the following equations for the
cooling Lorentz factor $\gamma_{\rm c}$ and Compton parameter $Y_{\rm
  c}$ at the cooling frequency:
\begin{eqnarray}
  \left(1+Y_{\rm c}\right)\gamma_{\rm c}&\,=\,&\gamma_{\rm c,syn}\ ,\nonumber\\
  Y_{\rm c}\left(1+Y_{\rm c}\right)&\,=\,&
  \frac{\epsilon_e}{\epsilon_{B-}}\left(
    \frac{\gamma_{\rm c}}{\gamma_{\rm min}}\right)^{2-p}\,\left[{\rm min}\left
      (1,\frac{\hat\gamma_{\rm c}}{\gamma_{\rm c}}\right)\right]^{(3-p)/2}\
  ,\label{eq:Ycgc}
\end{eqnarray}
with (see Nakar et al. 2009, Wang et al. 2010)
\begin{equation}
\gamma_{\rm c,syn}\,\equiv\,\left.\gamma_{\rm c}\right\vert_{Y_{\rm
    c}\,\rightarrow\,0}\ ,\quad
\hat\gamma_{\rm c}\,\equiv\,\frac{\gamma_{\rm b}m_ec ^2}{h\nu_{\rm
    c}(1+z)}\ .\label{eq:Ycgc2}
\end{equation}
We neglect more extreme cases in which the electron interacts with
low-frequency bands of the synchrotron spectrum, below $\nu_{\rm
  min}$. We then consider a synchrotron spectrum in the slow cooling
phase (generic in the cases that we study) $F_\nu\,\propto\,t_{\rm
  obs}^{-\alpha}\nu^{-\beta}$ with $\beta\,=\,3(p-1)/4$ above
$\nu_{\rm c}$ instead of $\beta\,=\,p/2$ when $\Upsilon_{\rm
  KN}\,<\,1$. We then verify a posteriori that the Compton parameter
in the X-ray range $Y_{\rm X}\,>\,1$, if the X-ray range is fitted
with this modified spectrum. In the GeV range, we always find
$Y_{>100\,{\rm MeV}}\,\ll\,1$ due to KN suppression; therefore, we
keep $\beta\,=\,p/2$ in that range. In Section~\ref{sec:dec}, we
incorporate the influence of decaying micro-turbulence, which modifies
further the time and frequency dependencies of the synchrotron
afterglow flux.

Finally, let us stress that while we assume that the bulk of the
emission at energy $>100\,{\rm MeV}$ originates from synchrotron
radiation, we do not exclude that a fraction of the highest energy
photons are actually produced by inverse Compton processes.  In
(homogeneous) small-scale turbulence, the high energy electrons suffer
only small angular deflections as they cross a coherence length of the
turbulence, so that their residence time (hence the acceleration time)
becomes substantially larger than the gyrotime, which sets the
residence time in a large-scale field (although advection impedes
acceleration in large-scale turbulence; see Lemoine, Pelletier \&
Revenu 2006). Hence, in small-scale turbulence peaked on a wavelength
$\lambda\,=\,10\,c/\omega_{\rm pi}$ with $\epsilon_{B+}\,=\,0.01$, the
maximum synchrotron photon energy falls to $1$-$3\,$GeV at an observed
time of $100\,$s for generic GRB afterglow parameters, see e.g. Kirk
\& Reville (2010), Bykov et al. (2012), Plotnikov, Pelletier \&
Lemoine (2013), Lemoine (2013) and Wang, Liu \& Lemoine (2013),
compared to a few tens of GeV for the ideal case of Bohm acceleration
on a gyrotime, e.g. Lyutikov (2010). In a decaying micro-turbulence,
the maximum photon energy does not depart much from the value for
homogeneous small-scale turbulence with $\epsilon_{B+}\,=\,0.01$, see
Lemoine (2013), because the highest energy electrons cool on a
relatively short time-scale, in regions of strong $\epsilon_B$ and at
the same time interact with modes of wavelength larger than
$\lambda$. For instance, a value of $\simeq\,2\,$GeV is derived at
$100\,$s assuming that the minimum scale of the turbulence
$\lambda=10\,c/\omega_{\rm pi}$ for a decay index $\alpha_t=-0.5$ and
a damping time of the turbulent modes $\tau\propto \lambda^2$.  In
this context, one should thus expect that photons of energy
$\gtrsim\,10\,$GeV do not originate from synchrotron radiation, but
from inverse Compton interactions, see Wang, Liu \& Lemoine (2013).  However,
the bulk of the emission $>100\,$MeV can be produced by synchrotron
radiation and we make this assumption in the present work. The energy
interval $100\,{\rm MeV} - 10\,{\rm GeV}$ indeed represents the bulk
of the emission for the GRBs seen with extended emission, because
their photon indices are $\simeq-2$, see Ackermann et al. (2013b).  Of
course, we verify a posteriori that the predicted synchrotron
self-Compton (SSC) contribution does not exceed the synchrotron flux
at energies $>100\,{\rm MeV}$.

\subsection{Application to four {\it Fermi}-LAT GRBs}
We now discuss the application of this exercise to four GRBs observed
in the radio, optical, X-ray and GeV range: GRB~090902B, GRB~090323,
GRB~090328 and GRB~110731A. We select them because four observational
constraints (corresponding to the four frequency bands) are required
to determine unambiguously the four parameters: $\epsilon_e$,
$\epsilon_{B-}$, $E$ and $n$. These four bursts have been discussed in
the literature: the first three by Cenko et al. (2011) and the last
one by Ackermann et al. (2013a). We will compare our results to these
studies in Sec.~\ref{sec:disc}.

\subsubsection{GRB~090902B}
We assume in the following $p=2.3$, as suggested by the previous
analyses of Cenko et al. (2011), Kumar \& Barniol-Duran (2010),
Barniol-Duran \& Kumar (2011) and Liu \& Wang (2011), and $k=0$
(constant density profile). The flux density at $100\,$MeV reads
\begin{equation}
  F_\nu\,\simeq\,6\times 10^{-9}\,{\rm
    Jy}\,E_{54}^{1.05}\epsilon_{B+,-2}^{0.05}\epsilon_{e,-1}^{1.2}t_2^{-1.15}\ ,
\end{equation}
so that its measured value $\simeq\,0.22\,\mu$Jy at a time
$t_{\rm obs}\,=\,50\,$s leads to
\begin{equation}
  E_{54}\,\simeq\,11.1\,\epsilon_{e,-1}^{-1.21}\
  .\label{eq:CGeV090902B}
\end{equation}
We have discarded the dependence on the Compton parameter
$Y_{>100\,{\rm MeV}}\,\ll\,1$ and on $\epsilon_{B+}$, since we assume
that the value of $\epsilon_{B+}$ that would enter this equation is
close to $0.01$, and its exponent is small.

For the optical range in the $R$~band at $\nu_{\rm opt}$, we assume
$\nu_{\rm min}\,<\,\nu_{\rm opt}\,<\,\nu_{\rm c}$ at $t_{\rm
  obs}\,=\,65\,000\,$s, with flux density $1.8\times
10^{-5}\,$Jy. Therefore, the optical flux
\begin{equation}
  F_\nu\,\simeq\,0.088\,{\rm
    Jy}\,E_{54}^{1.3}n_{-2}^{0.5}\epsilon_{B-,-2}^{0.8}
  \epsilon_{e,-1}^{1.2}t_2^{-0.9}
\end{equation}
leads to the constraint, once equation~\ref{eq:CGeV090902B} has been taken
into account:
\begin{equation}
\epsilon_{B-,-2}\,\simeq\,4.1\times
10^{-5}\,\epsilon_{e,-1}^{0.37}n_0^{-0.61}\ .\label{eq:Copt090902B}
\end{equation}
Quite interestingly, these two GeV and optical determinations lead by
themselves to very low values of $\epsilon_{B-}$, provided that
$\epsilon_{e,-1}$ and $n_0$ do not differ strongly from unity. The
radio flux at $\nu_{\rm rad}=8.5\,$GHz lies in the range of $\nu_{\rm
  rad}\,<\,\nu_{\rm min}\,<\,\nu_{\rm c}$ at $t_{\rm
  obs}\,\sim\,10^5\,$s, so that
\begin{equation}
  F_\nu\,\simeq\, 4.2\times10^{-5}\,{\rm
    Jy}\,E_{54}^{5/6}\epsilon_{B-,-2}^{1/3}n_{-2}^{1/2}t_{2}^{1/2}
  \epsilon_{e,-1}^{-2/3}\ ,
\end{equation}
to be matched to $F_\nu\,\sim\,1.3\times 10^{-4}\,$Jy at $4.8\times
10^5\,$s; when combined with the above
equations~(\ref{eq:CGeV090902B}) and (\ref{eq:Copt090902B}), this
implies
\begin{equation}
n_0\,\simeq\, 2.5\times 10^{-6}\,\epsilon_{e,-1}^{5.21}\ .\label{eq:Crad090902B}
\end{equation}
The decay rate in the X-ray range at $t_{\rm obs}\,>\,10^5\,$s
suggests that $\nu_{\rm c}<\nu$ (see Liu \& Wang 2011), which
therefore brings in complementary constraints relatively to the
optical and radio domains.  In principle, one should allow for a
different $\epsilon_B$ parameter in the region in which X-rays are
produced; here, we make however the approximation that this
$\epsilon_B\,\sim\,\epsilon_{B-}$.  In Section~\ref{sec:dec}, we
compute the afterglow allowing for the dependence of $\epsilon_B$ on
location, thus correcting this approximation.

If one first neglects KN suppression in the X-ray range, one is led to
a solution with $\epsilon_{e,-1}\,\sim\,2.7$, but with $\Upsilon_{\rm
  KN}\,\sim\,350$ at times $5\times 10^5\,$s, so that one needs to
include the KN suppression. Following the above algorithm, and using
the X-ray flux measurement between $0.3\,$ and $10\,$keV of
$2.2\times10^{-13}\,$erg/cm$^2$/s at $5.2\times 10^5\,$s, with
$\nu_{\rm X}>\nu_{\rm c}$, one derives $\epsilon_e$, hence the
parameter set
\begin{eqnarray}
  \epsilon_{e}&\,\simeq\,&0.46\ ,\quad
  E\,\simeq\,  1.8\times 10^{54}\,{\rm erg}\ ,\nonumber\\
  \epsilon_{B-}&\,\simeq\,&1.5\times 10^{-5}\ ,\quad
  n\,\simeq\, 7.0\times 10^{-3}\,{\rm cm}^{-3}\ .\label{eq:Sol090902B}
\end{eqnarray}
We also note that $\nu_{\rm c}\,\simeq\, 8.2\times 10^{16}\,{\rm Hz}$
at $5.2\times 10^5\,$s, $Y_{\rm c}\,\sim\,27$, just as $\nu_{\rm
  rad}\,<\,\nu_{\rm min}$ and $\nu_{\rm min}\,<\,\nu_{\rm
  opt}\,<\,\nu_{\rm c}$ at the respective times; the solution is
therefore consistent.

This light curve therefore indicates a low value for $\epsilon_{B-}$,
corresponding to a decay exponent
\begin{equation}
\alpha_t\,\simeq\, -0.44\pm0.10\ ,
\end{equation}
assuming $\epsilon_{B+}=0.01$ at $t\,=\,100\,\omega_{\rm pi}^{-1}$. We
used the value of $t_{\rm dyn}$ at time $10^5\,$s, at which the
predicted spectrum has been normalized to the optical and radio
data. We derive the uncertainty on $\alpha_t$ by propagating
conservative estimates of the uncertainties in the value of $p$, of
$k$ and the statistical errors of the data used for normalization. As
$p$ goes from $2.1$ to $2.5$, $\alpha_t$ changes from $-0.36$ to
$-0.48$. If $k=2$ instead of $0$, one finds
$\alpha_t=-0.51$~\footnote{The multi-wavelength light curve with a
  wind profile $k=2$ does not provide as good a fit to the data as
  that with $k=0$; however, it leads to a relatively high external
  wind parameter at early times, $A\,\sim\,10^{35}\,$cm$^{-1}$, which
  in turn implies a significant inverse Compton contribution at
  $>100\,{\rm MeV}$. Such a contribution could potentially explain the
  origin of the highest energy photon at $\,\sim\,30\,$GeV, which is
  difficult to account for in a scenario with $k=0$; see Wang, Liu \&
  Lemoine (2013).}. For this burst, scintillation in the radio range
provides the largest source of uncertainty, leading to a conservative
factor $\sim 3$ uncertainty on the flux, which in turn leads to an
error $\simeq\,0.03$ on $\alpha_t$. In total, we estimate the
uncertainty $\Delta\alpha_t\,\simeq\,0.10$.

\subsubsection{GRB~090323}
We repeat the same exercise with GRB~090323, which has been observed
at $>100\,$MeV up to a few hundred seconds, and in the X-ray, optical
and radio domains, short of a day onwards. In what follows, we use
$p=2.5$, slightly smaller than the value found by Cenko et al. (2011)
in their best fit, and $k=2$. The $>100\,$MeV flux is normalized to
$\phi(>100\,{\rm MeV})\,=\,1.5 \times
10^{-5}\,$photon~cm$^{-2}$~s$^{-1}$ at $350\,$s, leading to
\begin{equation}
E_{54}\,\simeq\,27.7\,\epsilon_{e,-1}^{-1.33}\ ,\label{eq:CGeV090323}
\end{equation}
while the optical flux is normalized to $1.3\times 10^{-5}\,$Jy at
$1.6\times 10^5\,$s, assuming $\nu_{\rm min}\,<\,\nu_{\rm
  opt}\,<\,\nu_{\rm c}$, leading to
\begin{equation}
  \epsilon _{B,-2}\,\simeq\,2.1\times 10^{-3}  A_{35}^{-1.14} 
\epsilon_{e,-1}^{-0.38}\ ,\label{eq:Copt090323}
\end{equation}
once equation~(\ref{eq:CGeV090323}) has been taken into account; then,
normalization to the radio flux $2.\times 10^{-4}$\,Jy at $4.3\times
10^5\,$s with $\nu_{\rm rad}\,<\,\nu_{\rm min}\,<\,\nu_{\rm c}$ leads
to
\begin{equation}
A_{35}\,\simeq\,0.98\,\epsilon_{e,-1}^2\ .\label{eq:Crad090323}
\end{equation}
Here as well, note that the radio, optical and GeV constraints lead to
a very low value for $\epsilon_B$, if one assumes a parameter
$\epsilon_e$ close to the value inferred in PIC simulations,
$\epsilon_{e,-1}\,\sim\, 1$. To account for the X-ray flux,
$\,\simeq\,10^{-13}\,$erg/cm$^2$/s at $2.5\times 10^5\,$s, it is here
as well necessary to consider the influence of KN suppression, which
eventually leads to
\begin{eqnarray}
  \epsilon_{e}&\,\simeq\,&0.25\ ,\quad
  E\,\simeq\,  8.1\times 10^{54}\,{\rm erg}\ ,\nonumber\\
  \epsilon_{B-}&\,\simeq\,&1.8\times 10^{-6}\ ,\quad
  A\,\simeq\, 6.1\times 10^{35}\,{\rm cm}^{-1}\ .\label{eq:Sol090323}
\end{eqnarray}
This corresponds to a decay index
\begin{equation}
\alpha_t\,\simeq\,-0.54\pm0.09  \ ,
\end{equation}
where the error accounts for a factor of 2 uncertainty on the GeV flux
(leading to $\pm0.06$ on $\alpha_t$), a factor of 2 uncertainty on the
radio determination (leading to $\pm 0.03$) and an uncertainty $\Delta
p=\pm0.2$ (leading to $\pm 0.04$); finally, if $k=0$ instead of $k=2$,
one finds $\alpha_t=-0.50$.

\subsubsection{GRB~090328}
The multi-wavelength light curve for this burst is rather similar to
that of GRB~090323, and we proceed analogously. Using a $>100\,$MeV
flux of $2.9\times 10^{-6}$\,photon~cm$^{-2}$~s$^{-1}$ at $1.1\times
10^3\,$s, we obtain
\begin{equation}
E_{54}\,\simeq\,2.1\,\epsilon_{e,-1}^{-1.33}\ ,\label{eq:CGeV090328}
\end{equation}
while the optical flux is normalized to $3\times 10^{-5}\,$Jy at
$0.6\times 10^5\,$s (with $\nu_{\rm min}\,<\,\nu_{\rm
  opt}\,<\,\nu_{\rm c}$), leading to
\begin{equation}
  \epsilon _{B,-2}\,\simeq\,1.5\times 10^{-3}  A_{35}^{-1.14} 
\epsilon_{e,-1}^{-0.38}\ .\label{eq:Copt090328}
\end{equation}
Normalization to the radio flux $6\times 10^{-4}$\,Jy at $3\times
10^5\,$s ($\nu_{\rm rad}\,<\,\nu_{\rm min}\,<\,\nu_{\rm c}$) leads to
\begin{equation}
A_{35}\,\simeq\,0.4\,\epsilon_{e,-1}^2\ .\label{eq:Crad090328}
\end{equation}
The X-ray flux is normalized to $2.7\times10^{-12}\,$erg/cm$^2$/s at
$0.63\times 10^5\,$s, in the KN regime, which leads to $\epsilon_e$,
hence
\begin{eqnarray}
  \epsilon_{e}&\,\simeq\,&0.19\ ,\quad
  E\,\simeq\,  0.88\times 10^{54}\,{\rm erg}\ ,\nonumber\\
  \epsilon_{B-}&\,\simeq\,&7.6\times 10^{-6}\ ,\quad
  A\,\simeq\, 1.5\times 10^{35}\,{\rm cm}^{-1}\ .\label{eq:Sol090328}
\end{eqnarray}
This corresponds to a decay index
\begin{equation}
\alpha_t\,\simeq\,-0.46\pm0.11  \ ,
\end{equation}
at time $10^5\,$s. The error accounts for a factor of 2 uncertainty on
the GeV flux (leading to $\pm0.06$ on $\alpha_t$), a factor of 2
uncertainty on the radio determination (leading to $\pm 0.02$) and an
uncertainty $\Delta p=\pm0.2$ (leading to $\pm 0.08$); finally, if
$k=0$ instead of $k=2$, one finds $\alpha_t=-0.42$.

\subsubsection{GRB~110731A}
This burst presents the most comprehensive multi-wavelength follow-up
of a LAT burst with extended emission at $>100\,$MeV; X-ray and
optical start short of $100\,$s, while $>100\,$MeV emission is still
ongoing. Unfortunately, there are no radio detections for this burst,
only an upper limit of $5\times 10^{-5}\,$Jy at $0.58\times 10^5\,$s
(Zauderer et al. 2011). Nevertheless, one can obtain strong
constraints on $\epsilon_B$, by noting that the optical frequency
$\nu_{\rm opt}=5.5\times 10^{14}\,$Hz must satisfy $\nu_{\rm
  opt}\,>\,\nu_{\rm min}$ at $t_{\rm obs}=100\,$s, because the optical
decays as a power law with index $\alpha\,\simeq\,1.37$; if the
opposite inequality were to hold at this time, one would rather
observe $\alpha=0$ for slow cooling, or $\alpha=1/4$ for fast
cooling. We thus write $\nu_{\rm min}\,=\,C_\nu\,\nu_{\rm opt}$ with
$C_\nu\,>\,1$ at $100\,$s, which imposes
\begin{equation}
  \epsilon _{B,-2}\,\simeq\, 5.1\times10^{-4}C_\nu^{-2}E_{54}^{-1}
\epsilon _{e,-1}^{-4}\ .\label{eq:Cnu110731A}
\end{equation}
Here and in the following, we assume $p=2.1$ and $k=2$.  Given that
$C_\nu>1$, this obviously restricts $\epsilon_B$ to very low values,
if $E$ and $\epsilon_e$ take close to standard values.  We next
normalize the predicted $F_\nu$ to the observed optical flux density
$\,\simeq\,3.5\times 10^{-4}\,$Jy at $1100\,$s, assuming $\nu_{\rm
  min}\,<\,\nu_{\rm opt}\,<\,\nu_{\rm c}$ (verified a posteriori), which
leads to
\begin{equation}
A_{35}\,\simeq\, 1.5\,C_\nu^{1.55} \epsilon_{e,-1}^2\ .\label{eq:Copt110731A}
\end{equation}
The above two conditions imply a radio flux which is a factor of
$\simeq4.1$ in excess of the observational upper bound; this remains
reasonable given the amount of scintillation typically expected at
this time, and seen in the other bursts. We then use the $>100\,$MeV
flux, $\phi(>100\,{\rm MeV})\,\simeq\,8.4\times
10^{-5}$\,photon~cm$^{-2}$~s$^{-1}$ at $26\,$s, to derive
\begin{equation}
E_{54}\,\simeq\,2.5\,\epsilon_{e,-1}^{-1.07}\ ,\label{eq:CGeV110731A}
\end{equation}
and finally the X-ray flux, $2\times 10^{-9}\,$erg/cm$^2$/s at
$100\,$s, assuming $\nu_{\rm c}<\nu_{\rm X}$. For this burst, KN
suppression is not effective at such an early time and it can be
neglected in the normalization; however, $\nu_{\rm c}$ is eventually
found to be close to $1\,$keV, which makes this solution only
approximate. In Section~\ref{sec:dec}, we derive a better fit by
adjusting by hand the missing parameter $\epsilon_{e,-1}$ under the
above constraints. Modulo this small uncertainty, the X-ray flux leads
to
\begin{eqnarray}
  \epsilon_{e}&\,\simeq\,&0.021\,C_\nu^{-0.50}\ ,\quad
  E\,\simeq\,  13.\times 10^{54}\,C_\nu^{0.54}\,{\rm erg}\ ,\nonumber\\
  \epsilon_{B-}&\,\simeq\,&1.9\times 10^{-4}\,C_\nu^{-0.53}\ ,\quad
  A\,\simeq\, 0.068\times 10^{35}\,C_\nu^{0.54}\,{\rm cm}^{-1}\
  .\nonumber\\
& & \label{eq:Sol110731A}
\end{eqnarray}
This implies a decay index
\begin{equation}
\alpha_t\,\simeq\,-0.35\frac{1+0.51\ln C_\nu}{1+0.10\ln C_\nu}\ ,
\end{equation}
at $t_{\rm obs}=1100\,$s. Assuming $C_\nu=1$, we estimate a
conservative uncertainty on $\alpha_t$ to be
$\Delta\alpha_t\,\simeq\,\pm0.2 $ given that a factor of $2$
uncertainty on the GeV flux leads to an error $\pm0.10$, $p=2.01$
leads to $\alpha_t=-0.14$ while $p=2.3$ leads to
$\alpha_t=-0.53$. Note that the light curves leave very little
ambiguity on the density profile (Ackermann et al. 2013); therefore,
we do not consider $k=0$.

\subsection{Multi-wavelength light curves in a decaying turbulence}\label{sec:dec}
We now include the effect of decaying micro-turbulence. The changing
magnetic field modifies the spectral shape of electrons with
$\gamma>\gamma_{\rm c}$, as well as the characteristic frequencies and
their evolution in time (Lemoine 2013). With respect to the previous
two-zone slow-cooling model, most of the difference concerns the X-ray
domain, which lies above $\nu_{\rm c}$. The spectrum is computed as
follows. 

At frequencies $<\nu_{\rm c}$, the standard synchrotron spectrum
holds, although the magnetic field value should be taken as the
partially decayed micro-turbulent value at the back of the blast, which
evolves in time:
\begin{equation}
  \delta B_{-}\,\simeq\, \delta B_{+}\,\left(t_{\rm dyn}/\tau_{\delta B}\right)^{\alpha_t/2}\,\propto\,
  \epsilon_{B+}^{1/2}\, t_{\rm obs}^{(\alpha_t-6)/8}\ .
\end{equation}
Of course, one recovers the standard time evolution in the limit
$\alpha_t\,\rightarrow\,0$.  

At frequencies $\nu_{\rm c}\,<\,\nu\,<\,\hat\nu_{\rm c}$, i.e. if
$\nu_{\rm c}\,<\,\hat\nu_{\rm c}$ ($\hat\nu_{\rm c}$ designing the
synchrotron peak frequency associated with $\hat\gamma_{\rm c}$), KN
suppression is ineffective, $\Upsilon_{\rm KN}(\nu)<1$; therefore, the
electrons cool in a uniform radiation background, but radiate their
synchrotron flux in a changing magnetic field, all along their cooling
history. This leads to a synchrotron spectral index
\begin{equation}
  \beta\,=\,\frac{p+\alpha_t/2}{2-\alpha_t/2}\quad
  \left[\nu\,>\nu_{\rm c},\ ,\Upsilon_{\rm KN}(\nu)\,<\,1\right]\
  , 
\end{equation}
see the Appendix of Lemoine (2013), Sec.~A3.

To account for the influence of KN suppressed inverse Compton losses
at frequencies $\nu\,>\,{\rm max}\left(\nu_{\rm c}\ ,\,\hat\nu_{\rm
    c}\right)$, we proceed as follows. We first solve for $\gamma_{\rm
  c}$ and $Y_{\rm c}$ as in equations~\ref{eq:Ycgc} and
\ref{eq:Ycgc2}, using however a value $\delta B_-$ for the magnetic
field at the back of the blast.  We then solve for the cooling history
$\gamma_e(t)$ of an electron with initial Lorentz factor (meaning at
the shock front, $t$ representing the comoving since acceleration at
the shock) $\gamma_{e,0}\,>\,\gamma_{\rm c}$, considering that if
$\Upsilon_{\rm KN}(\nu)\,>\,1$, this electron interacts with a
radiation field of energy density $Y(\gamma_e)\delta B_-^2/(8\pi)$,
characterized by the Lorentz factor dependent Compton parameter
$Y(\gamma_e)$ (e.g. Li \& Waxman 2006, Nakar et al. 2009, Wang et
al. 2010):
\begin{equation}
Y(\gamma_e)\,\simeq\,Y_{\rm c}\,\left(\frac{\gamma_e}{\gamma_{\rm
    c}}\right)^{(p-3)/2}\ ,
\end{equation}
assuming $\hat\gamma_{\rm c}\,<\,\gamma_{\rm c}\,<\,\gamma_e$. Here as
well, we can neglect extreme cases in which the electron interacts
with the low-frequency bands of the spectrum, below $\nu_{\rm
  min}$. Solving for the cooling history in this radiation field, one
determines a cooling time-scale $t_{\rm
  cool}(\gamma_{e,0})\,\simeq\,t_{\rm dyn}(\gamma_{e,0}/\gamma_{\rm
  c})^{-(p-1)/2}$, and $\gamma_e(t)\,\simeq\, \gamma_{\rm c}(t/t_{\rm
  dyn})^{-2/(p-1)}$ for $t\,\gg\,t_{\rm
  cool}(\gamma_{e,0})$. Following Lemoine (2013), we then calculate
the individual electron synchrotron contribution, by integrating the
synchrotron power $\propto\,\gamma_e^2(t) \delta B^2(t)$ over this
cooling history; then we evaluate the contribution of the electron
population by folding the latter result over the injection
distribution function of electron Lorentz factors. This leads to a
synchrotron spectral index
\begin{equation}
  \beta\,=\,\frac{3(p-1)}{4}\frac{1+\alpha_t /6}{1
  - \alpha_t(p-1)/8}\quad \left[\nu\,>\nu_{\rm c},\ ,\Upsilon_{\rm KN}(\nu)\,>\,1\right]\ ,
\end{equation}
which tends to $3(p-1)/4$ as it should when $\alpha_t\rightarrow 0$
(non-decaying turbulence). 

\begin{figure}
\includegraphics[bb=10 40 410 360, width=0.49\textwidth]{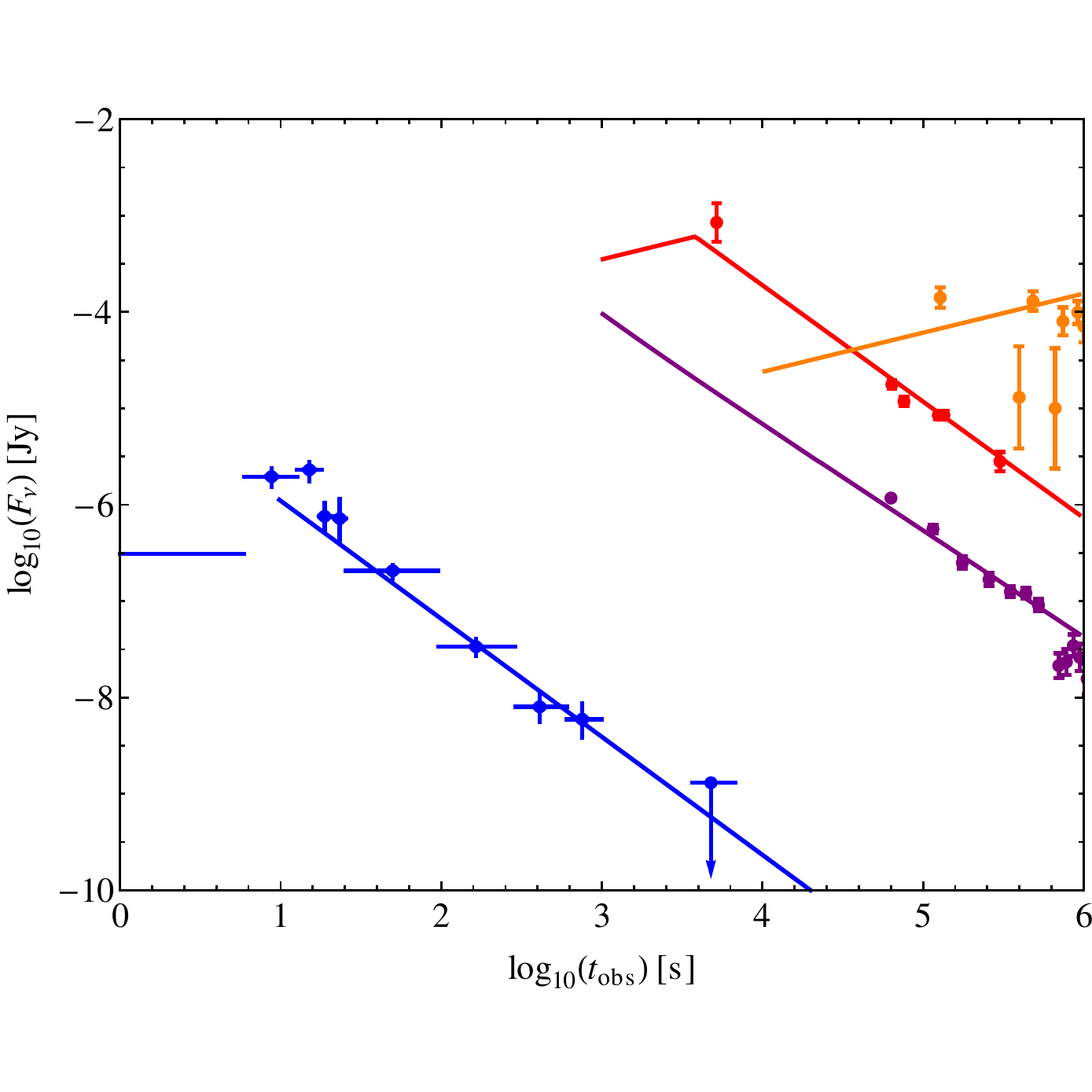}
\caption{Multi-wavelength light curve for GRB~090902B; orange: radio
  flux density, red: optical $R$-band flux density, purple: X-ray flux
  $\int_{0.3\,{\rm keV}/h}^{10\,{\rm keV}/h} F_\nu\,{\rm
    d}\nu/\left(2.4\times 10^{17}\,{\rm Hz}\right)$, blue: spectral
  flux density $F_\nu$ at $2.4\times 10^{22}\,$Hz; 
  parameter values: $E\,=\,1.6\times 10^{54}\,$ergs,
  $n=0.012\,$cm$^{-3}$, $\epsilon_e=0.50$, $p=2.3$, $k=0$ and
  $\alpha_t=-0.45$. Data taken from Cenko et al. (2011) and Abdo et
  al. (2009) and the {\it Swift} XRT repository data base (Evans et al. 2007,
  2009).
  \label{fig:090902B} }
\end{figure}

\begin{figure}
\includegraphics[bb=10 40 410 360, width=0.49\textwidth]{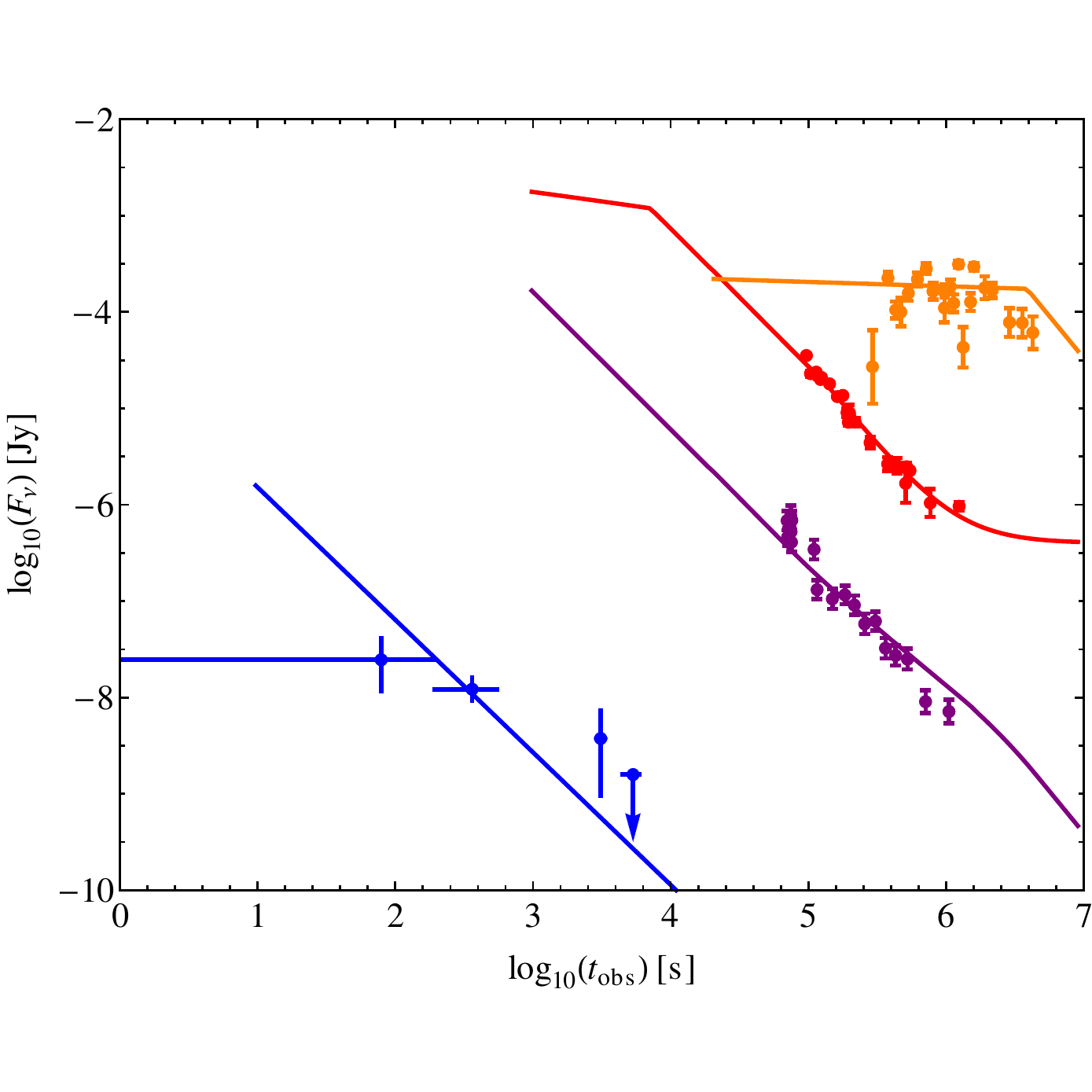}
\caption{Same as Fig.~\ref{fig:090902B}, for GRB~090323, except that the
  blue data indicates the $>100\,{\rm MeV}$ flux $\int_{100\,{\rm MeV}/h} F_\nu \,{\rm
    d}\ln\nu$. Parameter
    values: $E\,=\,5.4\times 10^{54}\,$ergs, $A=8.4\times
    10^{35}\,$cm$^{-1}$, $\epsilon_e=0.29$, $p=2.5$, $k=2$ and
    $\alpha_t=-0.54$. Data taken from Cenko et al. (2011), Piron et al.
    (2011) and the {\it Swift} XRT repository data base (Evans et al. 2007,
    2009); a constant $R$-band flux of $4\times 10^{-7}\,$Jy models the
    host galaxy emission.
    \label{fig:090323} }
\end{figure}

\begin{figure}
\includegraphics[bb=10 40 410 360, width=0.49\textwidth]{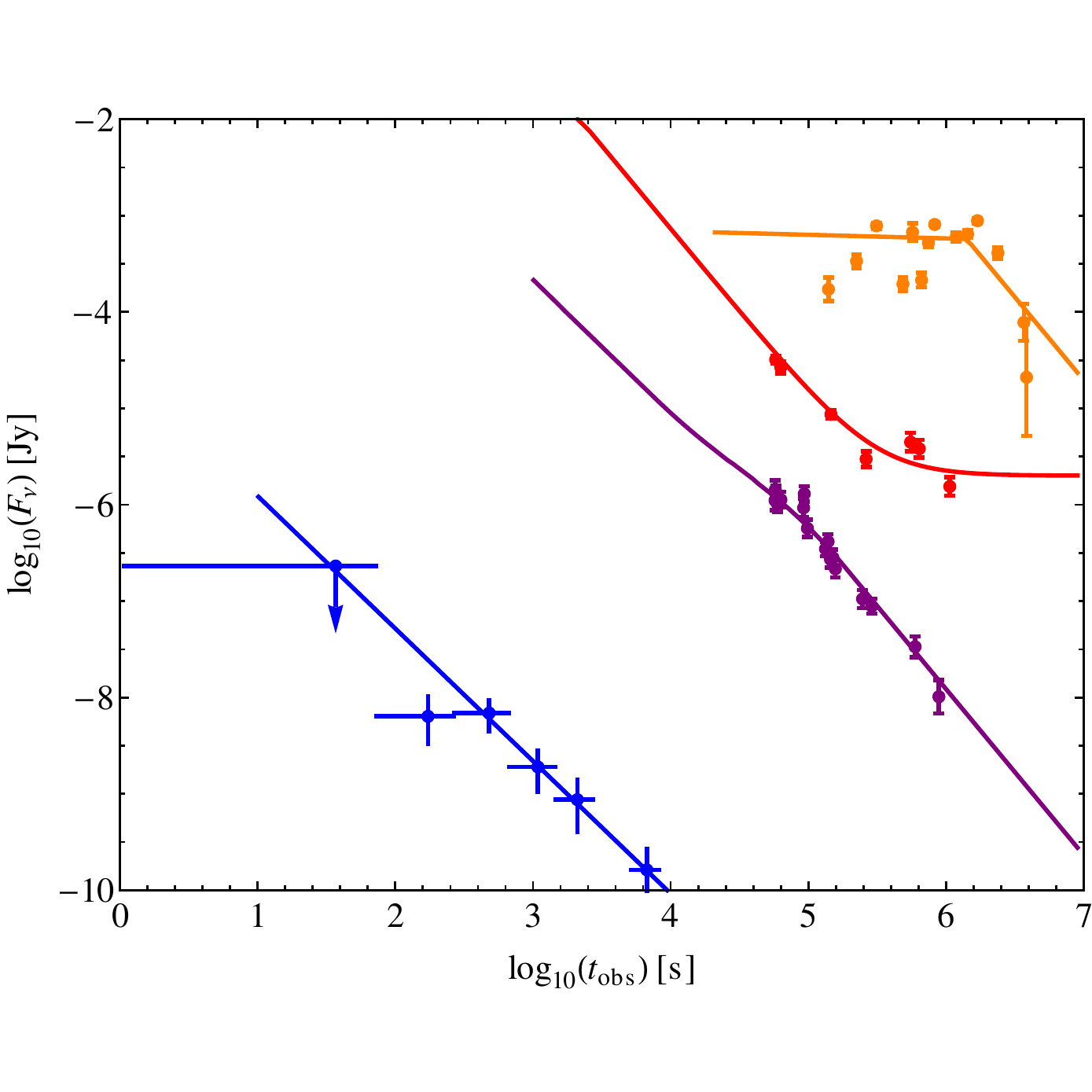}
\caption{Same as Fig.~\ref{fig:090323}, for GRB~090328 (using $U$-band
  optical data). Parameter values: $E\,=\,0.73\times 10^{54}\,$ergs,
  $A=1.5\times 10^{35}\,$cm$^{-1}$, $\epsilon_e=0.18$, $p=2.5$, $k=2$ and
  $\alpha_t=-0.45$. Data taken from Cenko et al. (2011), Piron et al. (2011)
  and the {\it Swift} XRT repository data base (Evans et al. 2007,
  2009);  a constant $R$-band flux of
  $2\times 10^{-6}\,$Jy models the host galaxy emission.
  \label{fig:090328} }
\end{figure}

\begin{figure}
  \includegraphics[bb=10 40 410 360, width = 0.49\textwidth]{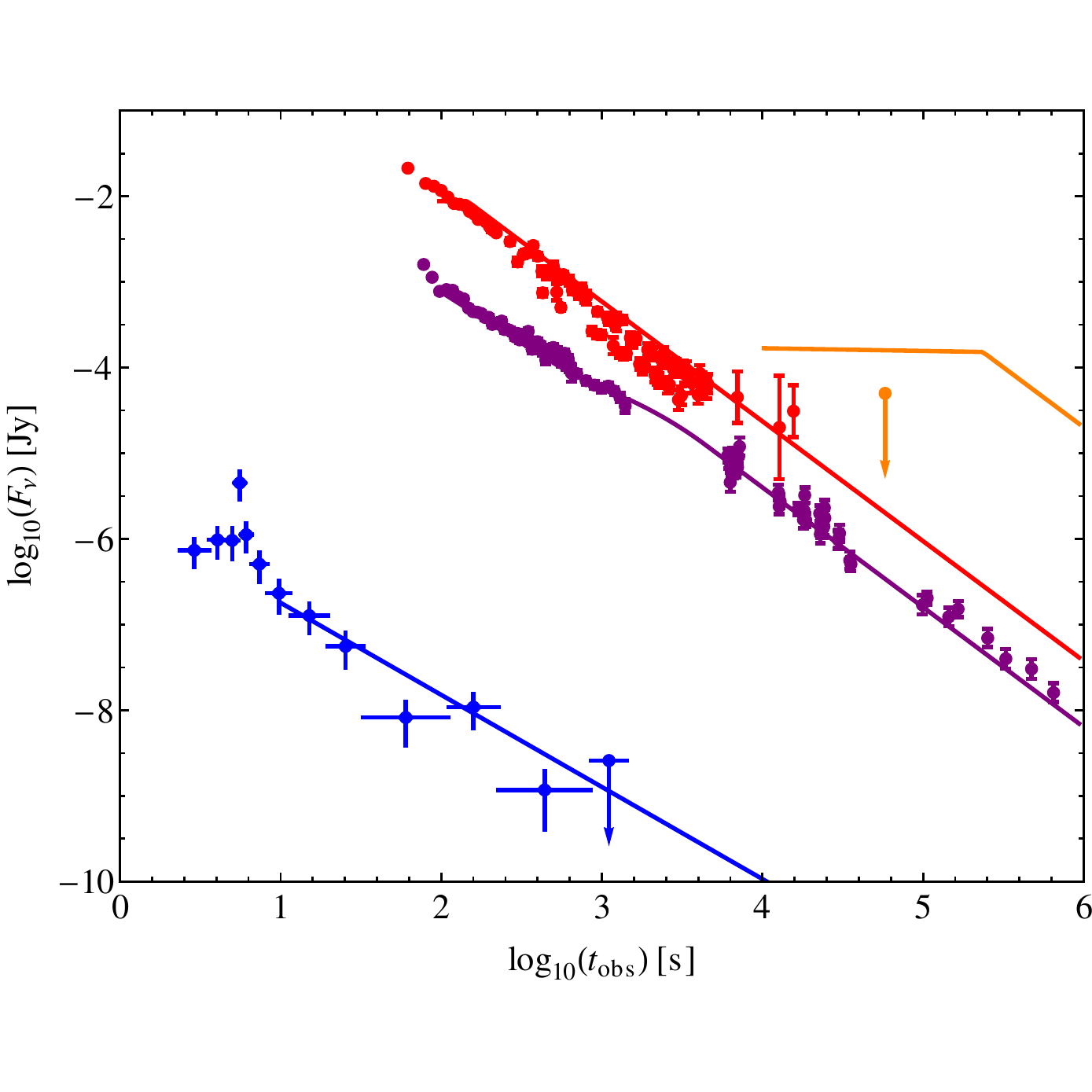}
\caption{Same as Fig.~\ref{fig:090323}, for GRB~110731A with in the optical
  range: UVOT $v$-band optical data, UVOT $w$-band data scaled to the
  $v$-band data, MoA $I$- and $V$-band data. Parameter values:
  $E\,=\,6.7\times 10^{54}\,$ergs, $A=0.15\times 10^{35}\,$cm$^{-1}$,
  $\epsilon_e=0.035$, $p=2.1$, $k=2$ and $\alpha_t=-0.38$. Data taken from
  Ackermann et al. (2013a) and the {\it Swift} XRT repository
  data base (Evans et al. 2007, 2009).
  \label{fig:110731A} }
\end{figure}

Finally, at $>100\,$MeV, we assume that inverse Compton losses are
negligible; hence, we use the above
$\beta=(p+\alpha_t/2)/(2-\alpha_t/2)$. This slight change of slope, as
compared to the two-zone determinations, implies slightly different
parameter values. The final estimates are given in the captions of
Figs.~\ref{fig:090902B},~\ref{fig:090323},~\ref{fig:090328} and
~\ref{fig:110731A}, which present the models of these multi-wavelength
light curves.

We have not attempted to obtain least-squares fits to these
multi-wavelength light curves, rather we have used the normalization
of the flux at several data points, as discussed in the previous
sections, derived the parameters, then plotted the predicted
multi-wavelength light curves. We have also neglected the possibility
of significant extinction in the optical domain, which could improve
the quality of the fit for GRB~110731A in particular. Moreover, our
numerical code computes the light curves for a decelerating blast
wave; it does not account for the initial ballistic stage, and neither
does it account for sideways expansion beyond jet break. We have
chosen to plot the $>100\,{\rm MeV}$ lightcurve assuming deceleration
of the blast beyond $10\,$s, which corresponds to initial Lorentz
factors $>700$ for GRB~110731A and GRB~090902B, for which $>100\,{\rm
  MeV}$ data exist at $10\,$s; one should note, however, that the
deceleration regime generally becomes valid beyond $T_{90}$, which
marks the duration of the prompt emission, and
$T_{90}\,=\,(150,\,70,\,25,\,8)\,$sec for GRB~090323, GRB~090328,
GRB~090902B and GRB~110731A respectively (Ackermann et al. 2013b).
Evidence for jet break is lacking in the four bursts, except possibly
for GRB~090902B (Cenko et al. 2011), in which case it would improve
the fit at times $\gtrsim\,10^6\,$s. Thus, there is room for improving
the quality of these fits, but it should not modify the value of
$\alpha_t$ derived in the previous sections beyond the quoted
uncertainties.

Finally, using the solutions indicated in the captions of the figures,
one can verify that synchrotron self-absorption effects are negligible
in the radio domain at the time at which the flux was normalized to
the data. One can also verify that for all bursts except GRB~090323,
the inverse Compton component provides a negligible contribution at
$>100\,{\rm MeV}$ at early times; for GRB~090323, this contribution is
a factor of 0.6 of the observed flux at $t_{\rm obs}\,=\,360\,$s, thus
non negligible. However, this remains within the error bars on the
flux normalization that we have adopted for this GRB, therefore we
neglect its influence. Future work should consider more detailed
multi-wavelength light curves including this inverse Compton
component, and possibly as well the effect of the maximal energy in
the $>100\,$MeV domain, as in Wang, Liu \& Lemoine (2013).

\section{Discussion}\label{sec:disc}

In the present work, we have argued that the afterglow of four GRBs
observed in the radio, optical, X-ray and at $>100\,$MeV by the {\it
  Fermi}-LAT instrument can be explained as synchrotron radiation in a
decaying micro-turbulence; such micro-turbulence and its decay through
collisionless phase mixing are expected on theoretical grounds, as
consequences of the formation of a relativistic collisionless shock in
a weakly magnetized environment. We have modelled the multi-wavelength
light curves of these four GRBs using first a simplified two-zone model
for the decaying turbulence, characterized in particular by
$\epsilon_{B-}$, which represents the value of $\epsilon_B$ at the
rear of the blast, where radio, optical and X-ray photons are
produced, and $\epsilon_{B+}\sim 0.01$, close to the shock where the
micro-turbulence has not yet had time to relax. Then we have used a
full synchrotron calculation assuming a power-law-decaying
micro-turbulence to improve on the above simplified model.

The low values of $\epsilon_{B-}$ that we derive here agree well with
the those derived by Kumar \& Barniol-Duran (2009, 2010),
Barniol-Duran \& Kumar (2011), He et al. (2011) and Liu \& Wang
(2011). There are however important differences in the interpretation
of these low values: Kumar \& Barniol-Duran (2009, 2010) argue that
all particles cool in the background shock compressed magnetic field
(including those producing $>100\,{\rm MeV}$ photons), which is
inferred of the order of $\sim\, 10\,\mu$G (upstream rest frame). We
rather argue that the particles cool in the post-shock decaying
micro-turbulence, which is self-generated in the shock precursor
through microinstabilities, and which actually builds up the
collisionless shock. As discussed in the introduction, this latter
interpretation is motivated by the large hierarchy between the
inferred values of $\epsilon_{B-}\,\sim\,10^{-6}$ to $10^{-4}$ and the much
smaller interstellar magnetization level $\sim\,10^{-9}$, indicating
that the background shock compressed field plays no role in shaping
the light curves. A power-law decay of the micro-turbulence behind the
shock front is also theoretically expected, e.g. Chang et al. (2008).
Furthermore, we provide a complete self-consistent model of the
synchrotron afterglow light curves in this scenario, based on and
improving the results of Lemoine (2013).  Within our interpretation,
we are thus able to constrain the value of the exponent of the
decaying micro-turbulence (assuming power-law decay), and we find a
consistent value among all bursts studied,
$-0.5\,\lesssim\,\alpha_t\,\lesssim\, -0.4$. This value turns out to
agree quite well with the results of the PIC simulations of Keshet et
al. (2009), see the discussion in Lemoine (2013).

These low values of $\epsilon_{B-}$ stand in stark contrast with other
determinations by Cenko et al. (2011) for GRB~090902B, GRB~090323 and
GRB~090328, and by Ackermann et al. (2013a) for GRB~110731A, who
systematically find values $\epsilon_B\,\sim\,0.01$. The key
difference turns out to come from the high-energy component
$>\,100\,$MeV. While in the present work, we assume that this extended
emission is synchrotron radiation from shock-accelerated electrons,
those studies do not incorporate the constraints from the high-energy
component.  Using the best-fitting models of Cenko et al. (2011) and
Ackermann et al. (2013), it is straightforward to calculate the ratio
$R_{>100\,{\rm MeV}}$ of the predicted photon flux $\phi(>100\,{\rm
  MeV})$ to the observed values\footnote{For GRB~090902B, we rather
  compare the spectral flux density at $2.4\times 10^{22}\,$Hz to the
  observed value.}:
\begin{eqnarray} {\rm 090902B}:&\,& R_{>100\,{\rm
      MeV}}\,\simeq\,7.2\times
  10^{-2}\quad\left(t_{\rm obs}\,=\,50\,{\rm s}\right)\nonumber\\
  {\rm 090323}:&\,& R_{>100\,{\rm MeV}}\,\simeq\,3.5\times
10^{-3}\quad\left(t_{\rm obs}\,=\,350\,{\rm s}\right)\nonumber\\
{\rm 090328}:&\,& R_{>100\,{\rm MeV}}\,\simeq\,1.9\times
10^{-2}\quad\left(t_{\rm obs}\,=\,1100\,{\rm s}\right)\nonumber\\
{\rm 110731A}:&\,& R_{>100\,{\rm MeV}}\,\simeq\,1.1\times
10^{-2}\quad\left(t_{\rm obs}\,=\,30\,{\rm s}\right)\
.\nonumber\\
& &  \label{eq:testsol}
\end{eqnarray}
The result is rather striking: those models do not explain the
high-energy component, in spite of the excellent quality of the fits
obtained in the other domains, e.g. Cenko et al. (2011). Ultimately,
this results from degeneracy in the parameter space, when only three
wavelength bands are used to determine the four parameters $E$, $n$,
$\epsilon_e$ and $\epsilon_B$ (assuming that some extra information is
available to determine $p$ and $k$, e.g. the time
behaviour). Specifically, the models of Cenko et al. (2011) and
Ackermann et al. (2013) present solutions that are degenerate up to
the choice of one of the above parameters, say $\epsilon_e$. To verify
this, one can explicitly repeat the above exercises, neglecting the
$>100\,{\rm MeV}$ data. By tuning $\epsilon_e$, one can then find
similar light curves, with different values of the parameters. These
different sets of solutions also correspond to different values of
$Y_{\rm c}$; the solutions of Cenko et al. (2011) and Ackermann et
al. (2013) systematically have $Y_{\rm c}\,\lesssim\,1$, while ours
rather corresponds to $Y_{\rm c}\,\gg\,1$. When $Y_{\rm c}\,>\,1$, the
solution scales differently with $\epsilon_e$, because of the
influence of inverse Compton losses in the X-ray domain
(notwithstanding possible KN suppression). As $Y_{\rm c}\,\gg\,1$, one
recovers our solutions up to the ambiguity in the choice of
$\epsilon_e$. This ambiguity is eventually raised by the normalization
to the $>100\,{\rm MeV}$ flux, leading to the present low $\epsilon_B$
values.

Going one step further, one should envisage the possibility that
earlier (pre-{\it Fermi}) determinations of the microphysical parameters
could be affected by a similar bias. The detailed analysis of
Panaitescu \& Kumar (2001, 2002) indicates indeed a broad range of
values of $\epsilon_{B-}$ for any GRB, spanning values from
$\sim10^{-6}$ up to $10^{-1}$. Thus, $\epsilon_{B-}$ is poorly
known. In very few cases, such as the famous GRB~970508, a synchrotron
self-absorption break seems to appear in the radio band. In these
cases, using the radio data in both optically thin and thick regimes,
as well as the optical and X-ray data, one has four bands for four
parameters, then all the parameters can be determined. A large value
for the magnetic field, $\epsilon_{B-}\sim0.01$, is obtained for
GRB~970508 by Wijers \& Galama (1999). However, the absorption break
in radio may not be clear given the bad quality of radio data (due to
strong scintillation). A recent re-analysis of GRB~970508 by Leventis
et al. (2013) also finds a variety of solutions, including one with a
low value of $\epsilon_{B-}$, when no ad hoc extra constraint is
imposed on the parameters. Future work should consider carefully the
uncertainty in the determination of $\epsilon_{B-}$ in such bursts.

Taken at face value, the present results suggest that the
magnetization of the blast can be described as the partial decay of
the micro-turbulence that is self-generated at the shock; it also
suggests that evidence for further amplification of this turbulence is
lacking, at least in the bursts observed by the {\it Fermi}-LAT
instrument.

While this paper was being completed, GRB130427A has been observed
with the {\it Fermi}-LAT instrument with unprecedented statistics,
with detailed follow-up observations in the radio, optical and X-ray
(see e.g. Laskar et al. 2013; Fan et al. 2013; Tam et al. 2013, and
references therein). As discussed in Tam et al. (2013), this GRB
presents strong evidence for the emergence of the SSC component at
high energies $\gtrsim 1-10\,{\rm GeV}$ above the synchrotron
component.  In a forthcoming paper, Liu, Wang \& Wu (2013) model the
multi-wavelength light curve of this GRB in a similar spirit to the
present analysis and derive in particular a value
$\epsilon_B\,=\,1.7\times 10^{-5}$. From their afterglow parameters,
one then infers $\alpha_t \,\simeq\,-0.44$, in excellent agreement
with the values derived here.

\bigskip

\noindent {\bf Acknowledgments:} We thank F. Piron for useful
discussions. This work has been supported in part by the PEPS/PTI
programme of the INP (CNRS), by the NSFC (11273005), the MOE
Ph.D. Programmes Foundation, China (20120001110064) and the CAS Open
Research Programme of Key Laboratory for the Structure and Evolution of
Celestial Objects, as well as the 973 Programme under grant
2009CB824800, the NSFC under grants 11273016, 10973008, and 11033002,
the Excellent Youth Foundation of Jiangsu Province (BK2012011). This
work made use of data supplied by the UK Swift Science Data Centre at
the University of Leicester.


\begin{thebibliography}{}

\bibitem[]{} Abdo, A. A. et al. (Fermi Collaboration), 2009, ApJ, 706, L138

\bibitem[]{} Ackermann et al. (Fermi Collaboration), 2013a, ApJ, 763, 71

\bibitem[]{} Ackermann et al. (Fermi Collaboration), 2013b,
  arXiv:1303.2908

\bibitem[]{} Barniol-Duran, R., Kumar, P., 2011, MNRAS, 417, 1584

\bibitem[]{} Bykov, A., Gehrels, N., Krawczynski, H., Lemoine, M.,
  Pelletier, G., Pohl, M., 2012, Space Sci. Rev., 173, 309

\bibitem[]{} Cenko, S. B. et al., 2011, ApJ, 732, 29

\bibitem[]{} Chang, P., Spitkovsky, A., Arons, J., 2008, ApJ, 674,
 378

\bibitem[]{} Evans, P. A. et al. (Swift-XRT), 2007, A\&A, 469, 379

\bibitem[]{} Evans, P. A. et al. (Swift-XRT), 2009, MNRAS, 397, 1177

\bibitem[]{} Fan, Y.-Z. et al., 2013, arXiv:1305.1261

\bibitem[]{} Gruzinov, A., Waxman, E., 1999, ApJ, 511, 852

\bibitem[]{} Haugb\o lle, T., 2011, ApJ, 739, 42

\bibitem[]{} He, H.-N., Wu, X.-F., Toma, K., Wang, X.-Y., M\'esz\'aros,
  P., 2011, ApJ, 733, 22

\bibitem[]{} Keshet, U., Katz, B., Spitkovsky, A., Waxman E., 2009,
  ApJ, 693, L127

\bibitem[]{} Kirk, J., Reville, B., 2010, ApJ, 710, 16

\bibitem[]{} Kumar, P., Barniol-Duran, R., 2009, MNRAS, 400, L75

\bibitem[]{} Kumar, P., Barniol-Duran, R., 2010, MNRAS, 409, 226

\bibitem[]{} Laskar, T. et al., 2013, arXiv:1305.2453

\bibitem[]{} Lemoine, M., Pelletier, G., Revenu, B., 2006, ApJ, 645, L129

\bibitem[]{} Lemoine, M., 2013, MNRAS, 428, 845

\bibitem[]{} Leventis, K., van der Horst, A. J., van Eerten, H. J.,
  Wijers, R. A. M. J., 2013, MNRAS, 431, 1026

\bibitem[]{} Li, Z., Waxman, E., 2006, ApJ, 651, L328

\bibitem[]{} Liu, R., Wang, X.-Y., 2011, ApJ, 730, 1

\bibitem[]{} Liu, R., Wang, X.-Y., Wu, X.-F., 2013, ApJ, 773, L20

\bibitem[]{} Lyutikov, M., 2010, MNRAS, 405, 1809

\bibitem[]{} Martins, S. F., Fonseca, R. A., Silva, L. O., Mori,
  W. B., 2009, ApJ, 695, L189

\bibitem[]{} Medvedev, M. V., Loeb, A., 1999, ApJ, 526, 697

\bibitem[]{} Medvedev, M. V., Trier Frederiksen, J., Haugboelle, T.,
  Nordlund, A., 2011, ApJ, 737, 55

\bibitem[]{} Nakar, E., Ando, S., Sari, R., 2009, ApJ, 703, 675

\bibitem[]{} Panaitescu, A., Kumar, P., 2001, ApJ, 554, 667

\bibitem[]{} Panaitescu, A., Kumar, P., 2002, ApJ, 571, 779

\bibitem[]{} Piran, T, 2004, Rev. Mod. Phys., 76, 1143

\bibitem[]{} Piron F., McEnery J., Vasileiou V. and the Fermi-LAT and GBM Collaborations, 2011, AIP Conf. Proc. 1358, pp. 47

\bibitem[]{} Plotnikov, I., Pelletier, G., Lemoine, M., 2013, MNRAS,
  430, 1208

\bibitem[]{} Sari, R., Piran, T., Narayan, R., 1998, ApJ, 497, L17

\bibitem[]{} Sari, R., Esin,  A. A., 2001 ApJ, 548, 787

\bibitem[]{} Sironi, L., Spitkovski, A., 2011, ApJ, 726, 75

\bibitem[]{} Sironi, L., Spitkovski, A., 2013, ApJ, 771, 54

\bibitem[]{} Spitkovsky, A., 2008, ApJ 682, L5

\bibitem[]{} Tam, P.-H. T., Tang, Q.-W., Hou, S.-J., Liu, R.-Y., Wang,
  X.-Y., 2013, ApJ, 771, L13

\bibitem[]{} Wang, X.-Y., He, H.-N., Li, Z., Wu, X.-F., Dai, X.-G.,
  2010, ApJ, 712, 1232

\bibitem[]{} Wang, X.-Y., Liu, R., Lemoine, M., 2013, ApJ, 771, L33

\bibitem[]{} Wijers, R. A. M. J., Galama, T. J., 1999, ApJ, 523, 177

\bibitem[]{} Zauderer, A., Berger, E., Frail, D. A. et al., 2011, GRB
  Coordinates Network, Circular Service, 12227, 1
\end{thebibliography}
\end{document}